# Compressed radiotherapy treatment planning (CompressRTP): A new paradigm for rapid and high-quality treatment planning optimization

code and sample data available at https://github.com/PortPy-Project/CompressRTP

**Mojtaba Tefagh[1], Gourav Jhanwar[2], Masoud Zarepisheh[2*]**

[1]Edinburgh Futures Institute, University of Edinburgh, Edinburgh, Scotland

[2]Department of Medical Physics, Memorial Sloan Kettering Cancer Center, New York, NY

[*]Email address: zarepism@mskcc.org

**Background:** Radiotherapy treatment planning involves solving large-scale optimization problems that are often approximated and solved sub-optimally due to time constraints. Central to these problems is the dose influence matrix—also known as the $dij$ matrix or dose deposition matrix—which quantifies the radiation dose delivered from each beamlet to each voxel. Our findings demonstrate that this matrix is highly compressible, enabling a compact representation of the optimization problems and allowing them to be solved more efficiently and accurately.

**Purpose:** To develop a compressed radiotherapy treatment planning framework based on a sparse-plus-low-rank matrix compression technique. This approach circumvents conventional sparsification methods that discard small matrix elements—often representing scattering components—and may compromise the quality of the treatment plan.

**Methods:** We precompute the primary ($S$) and scattering ($L$) dose contributions of the dose influence matrix $A$ separately for photon therapy, expressed as: $A = S + L$. Our analysis reveals that the singular values of the scattering matrix $L$ exhibit exponential decay, indicating that $L$ is a low-rank matrix. This allows us to compress $L$ into two smaller matrices: $L^{m \times n} = H^{m \times r} W^{r \times n}$, where $r$ is relatively small (approximately 5 to 10). Since the primary dose matrix $S$ is sparse, this supports the use of the well-established "sparse-plus-low-rank" decomposition technique for the influence matrix $A$, approximated as: $A \approx S + H \times W$. We introduce an efficient algorithm for sparse-plus-low-rank matrix decomposition, even without direct access to the scattering matrix. This algorithm is applied to optimize treatment plans for ten lung and ten prostate patients, using both compressed and sparsified versions of matrix $A$. We then evaluate the dose discrepancy between the optimized and final plans. We also integrate this compression technique with our in-house automated planning system, ECHO, and evaluate the dosimetric quality of the generated plans with and without compression.

**Results:** CompressRTP offers superior trade-offs between accuracy and computational efficiency, adjustable through algorithm parameters. For example, we achieved average reductions in dose discrepancy and optimization time of 73% and 20%, respectively, for ten prostate patients, and 83% and 13% for lung patients. By using the compressed matrix within our automated ECHO planning system, we maintained comparable PTV coverage while significantly enhancing the sparing of organs at risk. Specifically, mean doses to the bladder and rectum for prostate patients were reduced by 8.8% and 12.5%, respectively. For lung patients, mean doses to the lungs (left and right, excluding GTV) and heart were reduced by 10.8% and 11.2%, respectively, compared to plans generated with the sparsified matrix.

**Conclusion:** The proposed CompressRTP framework enables rapid, high-quality treatment planning without compromising data integrity and plan quality. By integrating CompressRTP with recent advancements in AI-driven influence matrix calculations, this platform has the potential to facilitate fast and efficient online adaptive radiotherapy treatment planning, enhancing both speed and accuracy in clinical workflows.

# 1. INTRODUCTION

Radiotherapy treatment planning is a complex decision-making process that aims to maximize tumor control while minimizing the risk of radiotoxic complications in nearby organs-at-risk (OARs). The patient's unique anatomy and the capabilities of the treatment machine define the feasible treatment options. Treatment planning algorithms navigate this search space to select a plan with the most favorable trade-offs between tumor eradication and normal tissue preservation. As schematically illustrated in Fig. 1, blue circles represent various treatment options with differing trade-offs, while red arrows indicate the algorithm's pathway toward identifying an optimal plan. A variety of planning algorithms are available today, ranging from classical techniques such as conventional trial-and-error, knowledge-based methods[1,2], prioritized optimization[3–8], and multi-criteria optimization (MCO)[9,10], to modern AI-based approaches like deep learning and reinforcement learning[11–15]. While the choice of treatment planning algorithm influences the pathway to the optimal plan, the actual treatment options

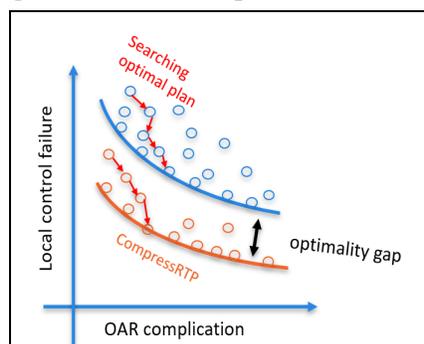

**Fig. 1:** A schematic illustrating that treatment planning algorithms explore feasible treatment options (blue circles) to identify a plan with optimal trade-offs. CompressRTP enhances this process by introducing improved options (orange circles), generated using precise dosimetry, helping to close the optimality gap in current practices.

are determined by the patient's anatomy and the fundamental principles of radiation physics.

During treatment planning optimization, available treatment options are calculated based on precomputed dosimetric data, typically represented in a large matrix known as the dose influence matrix (also referred to as the dose deposition matrix or the $dij$ matrix or dose deposition coefficients)[16,17]. This matrix quantifies the radiation dose delivered from each beamlet to every voxel in the patient. However, due to the complexity of processing such large and dense dosimetric data, these matrices are often truncated and sparsified to enhance computational efficiency—for example, by excluding small scattered radiation dose components. Using inaccurate dosimetric data during treatment planning can lead to the generation of suboptimal treatment options (represented by blue circles in Fig. 1) due to the "garbage-in, garbage-out" phenomenon. In other words, inaccuracies in the influence matrix result in significant discrepancies between the planned radiation dose and the final accurate dose used for evaluation. Currently, this issue is partially addressed through a "correction loop"[16,17], which intermittently integrates more accurate dose calculations into the optimization process. Although this heuristic method can correct minor discrepancies, it does not fully resolve the issue, leaving some level of plan suboptimality[18]. Additionally, in the case of constrained optimization—where clinical criteria are strictly enforced using hard constraints—each correction step requires solving the optimization problem from scratch, substantially increasing computational time due to the numerous correction steps often needed. This is because the interior point method[19–21], considered the state-of-the-art algorithm for solving constrained problems, is not "warm-start friendly" (i.e., it cannot fully leverage the previous solution).

Recent advancements in hardware and algorithm design have significantly accelerated accurate physics-based dose calculations (e.g., Acuros XB[22], Anisotropic Analytical Algorithm (AAA)[23], Collapsed Cone Convolution (CCC)[24]) and Monte Carlo[25–27] simulation-based methods. These developments have reached new heights with the advent of AI and deep learning, leading to various AI-based techniques capable of rapid (millisecond-scale) and precise dose calculations using patient CT scans and radiotherapy machine beam parameters as inputs[28–31]. However, current research has primarily focused on employing these fast and accurate dose calculations for final plan evaluation or plan quality assurance, rather than integrating them into treatment planning optimization. The primary challenge in incorporating these advanced dose calculations into treatment planning optimization is managing the large and dense dose influence matrices, which include detailed radiation scattering components. For context, current dose influence matrices used in optimization are approximately 95%-98% sparse (i.e., with only 2%-5% nonzero elements), whereas accurate, detailed matrices are entirely dense. This increase in density demands about 20-50 times more memory and significantly slows down optimization processes. A notable example of this problem is our in-house automated treatment planning system called ECHO[3,4,32–34] (Expedited Constrained Hierarchical Optimization). ECHO is integrated with Eclipse through its APIs and is actively used in our daily clinical practice, having treated over 10,000 patients to date. To manage computational efficiency, ECHO uses Eclipse's AAA dose calculation for computing the influence matrix but truncates the resultant dense matrix by ignoring small-value elements—typically any value less than 1% of the maximum element in the matrix.

In this paper, we demonstrate that the dose influence matrix is highly structured and amenable to a matrix decomposition known as sparse-plus-low-rank decomposition (also referred to as low-rank-plus-sparse decomposition). Unlike classical data compression techniques (e.g., ZIP, JPEG), sparse-plus-low-rank does not require data decompression and is ideal for matrix-vector operations in iterative optimization algorithms. Sparse-plus-low-rank representations have been widely used in fields such as computer vision, medical imaging, and statistics for fundamental tasks including foreground-background image separation, rapid image reconstruction, and principal component analysis of noisy data[2,35–38]. However, to the best of our knowledge, the application of this technique as a data compression method for improving computational efficiency is unprecedented—even outside the field of radiotherapy. We illustrate the applicability of this technique on a typical intensity-modulated radiation therapy (IMRT) optimization problem, for which we have made the code and sample data publicly available on [GitHub](). Additionally, we demonstrate the benefits of this technique by integrating it with our in-house automated planning system, ECHO.

## 2. METHODS

### 2.1 Compressed Radiotherapy Treatment Planning (CompressRTP)

IMRT treatment planning optimization can generally be formulated as the following optimization problem:

$$\min_x f(Ax, x) \ s.t. \ g(Ax, x) \leq 0, x \in X, \quad Problem \ (1)$$

where x represents the beamlet intensities, $A$ is the patient-specific dose influence matrix, and $Ax$ is the radiation dose delivered to the patient's body. The function $f$ is the objective function, and $g$ represents the constraints. The specific treatment planning technique employed—such as hierarchical optimization, MCO, or AI-based methods—determines the particular problems and parameters to be solved. Matrix $A$ is large, typically with 100,000 to 1,000,000 rows corresponding to the patient's voxels and 1,000 to 10,000 columns corresponding to the machine's beamlets. This matrix is the main source of computational bottleneck in solving the optimization problems. $A$ can be decomposed into $A = S + L$, where $S$ is a sparse matrix containing the primary dose contributions, and $L$ includes the scattering dose contributions. The rows and columns of matrix $A$ exhibit high correlations due to the spatial relationships between adjacent voxels and beamlets. These correlations result in a low-rank and compressible matrix, which can be mathematically verified by observing an exponential decay in the singular values of $A$ (illustrated by the blue line in Fig. 1a). The scattering matrix $L$ is even more compressible, evidenced by a sharper exponential decay in its singular values (red line in Fig. 1a). This suggests the use of sparse-plus-low-rank compression, as shown schematically in Fig. 1b.

The low-rank nature of $L$ allows it to be compressed into two simpler matrices through low-rank decomposition. Specifically, we can express $L$ (an $m$ by $n$ matrix) as the product of two matrices $H$ and $W$, i.e. $L = H \times W$. Here, $H$ is an $m$ by $r$ matrix—a "tall-skinny" matrix with many rows but relatively few columns—and $W$ is an $r$ by n matrix—a "wide-short" matrix with relatively few rows and many columns. The rank $r$ is relatively small (approximately 5 to 10). This low-rank decomposition is also the premise behind the well-known and widely used statistical tool called Principal Component Analysis (PCA). In this context, $H$ and $W$ can be conceptualized as containing the principal components of the columns and rows of matrix $L$, respectively.

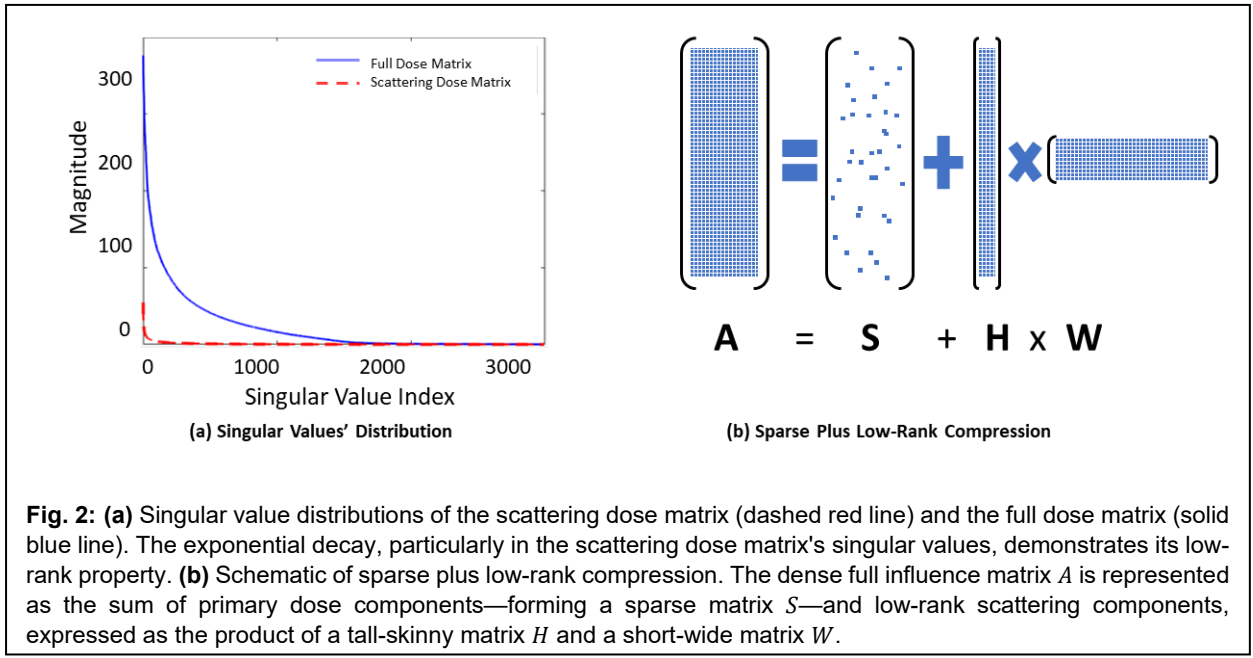

**Fig. 2: (a)** Singular value distributions of the scattering dose matrix (dashed red line) and the full dose matrix (solid blue line). The exponential decay, particularly in the scattering dose matrix's singular values, demonstrates its low-rank property. **(b)** Schematic of sparse plus low-rank compression. The dense full influence matrix $A$ is represented as the sum of primary dose components—forming a sparse matrix $S$—and low-rank scattering components, expressed as the product of a tall-skinny matrix $H$ and a short-wide matrix $W$.

In Fig. 2, we used the CERR open-source package[39] to obtain the influence matrix for a lung patient, where the primary and scattering dose contributions are provided separately. However, in scenarios where the dose calculation engine does not output the primary and scattering contributions individually—due to inherent algorithm design, computational overhead, or restricted access when the dose engine functions as a black box—it is still possible to decompose the matrix $A$ into $A = S + L$. In this decomposition, $S$ is a sparse matrix containing the large-magnitude elements (e.g., elements exceeding 1% of the maximum value of $A$), and $L$ is a dense, low-rank matrix comprising the remaining small-value elements. In our experiments, we employed the Eclipse API to calculate the influence matrix using the AAA[23] for dose calculation. In this context, the dose engine operates as a black box, and the Eclipse API does not provide separate outputs for the primary and scattering dose components. Despite this limitation, as demonstrated in Fig. A in the Appendix, the matrix of small-value elements $L$ still exhibits a very low-rank structure. Algorithm 1 presents the pseudocode for performing sparse-plus-low-rank compression of the influence matrix.

Algorithm 1 is straightforward and can be implemented with just a few lines of code in high-level programming languages like Python or MATLAB, primarily leveraging the readily available Singular Value Decomposition (SVD) technique. The algorithm has two hyperparameters: the sparsification threshold $\eta$ and the rank $r$ for the low-rank matrix decomposition. The process begins by assigning all elements greater than $\eta$ times the maximum value of $A$ (i.e., elements where $A_{ij} > \eta \times \max(A)$) to the sparse matrix $S$. Subsequently, SVD is performed on the residual matrix $L = A - S$, which contains the smaller-value elements. The decomposition outputs are then used to form the matrices $H$ (a tall, skinny matrix) and $W$ (a wide, short matrix), following the relation $L = H \times W$. In current practice, the small-value dose influence matrix $L$ is often omitted in treatment planning for computational efficiency, leading to the approximation $A \approx S$. However, our research demonstrates that the compressed representation $A \approx S + H \times W$ provides a more efficient and accurate representation of the data.

---

**Algorithm 1.** A pseudocode for sparse-plus-low-rank compression of the influence matrix
**Inputs:** Influence matrix $A^{m \times n}$, sparsification threshold $\eta$, rank $r$
**Outputs:** Sparse matrix $S^{m \times n}$, tall-skinny matrix $H^{m \times r}$, wide-short matrix $W^{r \times n}$ ($A \approx S + HW$)
$S \leftarrow 0$     /* a zero matrix with the same size as $A$ */
$S \leftarrow$ elements of A larger than $\eta \times max(A)$   /* retain large elements of A */
$L \leftarrow A - S$
$[U, \Sigma, V] = svd(L, r)$   /* Compute the truncated SVD of L, keeping the top r singular values $\Sigma$ and corresponding right/left vectors U / V */
$H \leftarrow U \times \Sigma$
$W \leftarrow V^T$

---

## 2.2 Treatment Planning Optimization Frameworks

In this study, we use the following commonly employed IMRT optimization problem, which can be considered a special case of Problem (1):

$$\min_{x} \sum_{s=1,\dots,N^S} (w_+^s \times \max(A^s x - p^s, 0)^2 + w_-^s \times \max(p^s - A^s x, 0)^2) + \lambda ||Px||_2^2$$

$$s.t.$$

$$\max(A^s x) \leq d_s^{max}, \quad s \in \text{Structures with max dose constraints,}$$
$$\text{mean}(A^s x) \leq d_s^{mean}, \quad s \in \text{Structures with mean dose constraints,}$$
$$x \geq 0, \quad \text{Non-negativity constraint,} \quad\quad Problem\ (2)$$

where $A^s$ denotes the rows of the influence matrix $A$ corresponding to voxels within structure s (s= 1,…,$N^S$). The term $p^s$ represents the prescribed dose, set to zero for organs-at-risk (OARs). The weights $w_+^s$ and $w_-^s$ penalize overdose and underdose violations, respectively ($w_-^s = 0$ for OARs). The matrix $P$ is a total-variation matrix aimed at minimizing variations across neighboring beamlets to enhance the smoothness of the fluence map, improving delivery efficiency. The parameter $\lambda$ scales the importance of fluence smoothing. The hard constraints enforce maximum and mean dose clinical criteria. Upon determining the optimal intensity $x$, a leaf sequencing process is carried out to finalize the optimal leaf motions.

In Section 3.1, we present our experiments on lung and prostate patients using Problem (2). The maximum and mean dose constraints are detailed in Table 1. The weight hyperparameters $w$ and $\lambda$ are manually adjusted to derive reasonable treatment plans. It is important to note that these parameters minimally influence our analysis of the compression effects. For lung patients, all relevant materials—including data, code, problem formulations, and hyperparameters—are available on our GitHub page. In these experiments, we solve Problem (2) by substituting the original dense matrix $A$ with either a sparse-only matrix $S$ or a compressed sparse-plus-low-rank matrix $S + HW$, derived from Algorithm 1 using various sparsification thresholds $\eta$ and ranks $r$. Due to the dependency of the optimal solution on the input matrix, it would be logical to denote the optimal solutions as $x^S$ and $x^{(S,H,W)}$. However, for notational simplicity, we slightly abuse the notation and refer to the optimal solution simply as $x$, assuming the context makes this clear. It is also worth mentioning that solving Problem (2) with the original matrix $A$ is often impractical due to memory limitations, system freezes, or excessively slow computations. For each configuration of the influence matrix, we solve Problem (2) and assess two main outcomes: accuracy—gauged by the dose discrepancies between the optimized dose and the final dose ($\Delta = Ax - Sx$ and $\Delta = Ax - (Sx + HWx)$)—and computational performance.

In Section 3.2, we investigate the effects of the existing "correction loop" technique both with and without the application of compression. We begin by solving Problem (2) using both compressed and non-compressed matrices and calculate the resulting dose discrepancy $\Delta$. To mitigate these discrepancies, we then solve Problem (2) again, this time incorporating $\Delta$ into the optimization. This adjustment involves updating the terms in the optimization Problem (2) from $Sx$ to $Sx + \Delta$ and from $Sx + HWx$ to $Sx + HWx + \Delta$.

In Section 3.3, we assess the impact of compression on the quality of the final treatment plans generated by ECHO, our in-house automated planning system. While a high-level overview of ECHO is provided here, detailed information can be found in previous publications[3,4,32–34,40,41]. ECHO solves two optimization problems sequentially, known as Step-1 and Step-2. Step-1 closely resembles Problem (2) but differs by setting the weights for the organs-at-risk (OARs) in the objective function to zero, focusing solely on optimizing the planning target volume (PTV) coverage and homogeneity. Step-2, also similar to Problem (2), sets the weights for the PTV to zero in the objective function and introduces additional constraints to maintain the results achieved in Step-1. In scenarios involving dose-volume constraints (DVHs), an additional preliminary step called Step-0 is implemented before Step-1. Step-0 is similar to Step-1 but includes extra linear constraints to approximate the DVH constraints[33]. After completing Step-0, low-dose voxels are identified based on the specified DVH constraints, and appropriate maximum dose constraints are incorporated into both Step-1 and Step-2 to ensure the satisfaction of the DVH constraints.

**Table 1.** Clinical dose constraints for lung and prostate.

| Prostate | | | Lung | | |
|---|---|---|---|---|---|
| Structure | Type | Constraint | Structure | Type | Constraint |
| PTV | Max | 110% | PTV | Max | 115% |
| Bladder | Max | 106% | Esophagus | Max | 110% |
|  | Mean | 66% |  | Mean | 34Gy |
| Rectum | Max | 106% |  | V(60Gy) | 17% |
|  | Mean | 44% | Heart | Max | 110% |
| Large Bowel | Max | 26.5Gy |  | Mean | 20Gy |
| Small Bowel | Max | 25Gy | Cord | Max | 50Gy |
| Femoral Heads | Max | 20Gy | Lungs (Excluding GTV) | Max | 110% |
| Urethra | Max | 100% |  | Mean | 21Gy |
|  | D(1cc) | 95% |  | V(20Gy) | 37% |

**2.3 Data Preparation and Computational Platform**

We conducted our study using data from ten prostate cancer patients and ten lung cancer patients. An expert physicist selected the beam configurations for each patient. We precomputed the dose influence matrix $A$ using the Eclipse™ API 16.1 (Varian Medical Systems, Inc., Palo Alto, California), utilizing the AAA version 16.1.0[23]. Table 2 summarizes the patient data. For each patient, the dimensions of the dose influence matrix $A$ are represented as the number of voxels by the number of beamlets.

We implemented the algorithm in Python and ran it on a PC with a 2.4 GHz Intel Xeon CPU and 256 GB of RAM. To access the data, obtain the optimal fluence in an Eclipse-compatible format, and solve the convex constrained optimization problems, we utilized the open-source platform PortPy[42] (Planning and Optimization for Radiation Therapy in Python), which leverages the CVXPy toolbox[43] alongside the MOSEK optimization engine[44]. For the results in Section 3.3, the optimized beamlet intensity values ($x$) were imported back into the Eclipse system for final leaf sequencing and dose calculation.

**Table 2.** Data summary of patient cases.

| Tumor type | PTV size (cc) | # of beams | # of beamlets | # of voxels | Prescription (p) |
|---|---|---|---|---|---|
| Prostate | 100-168 | 9 | 2978-4939 | 754176-1090129 | 25 Gy in 5 fractions |
| Lung | 87-737 | 7 | 1666-6650 | 317307-609761 | 60 Gy in 30 fractions |

## 3. RESULTS

### 3.1 Compression Reduces the Dose Discrepancy

In IMRT treatment planning optimization, two primary factors contribute to dose discrepancies between the optimized dose and the final dose: (1) the approximation of the dose influence matrix—specifically, using the truncated sparse matrix $S$ in optimization instead of the full original matrix $A$—and (2) the effects of leaf sequencing. Fig. 3 illustrates the DVH comparisons, showing the optimized dose $Sx$ (dotted lines), the full dose $Ax$ (solid lines), and the final dose after importing the optimal fluence $x$ into Eclipse and performing leaf sequencing (dashed lines). For consistency in comparisons, the plans are normalized so that $V_{100\%}(PTV) = 90\%$ for the prostate case and $V_{100\%}(PTV) = 80\%$ for the lung case. Fig. 3 highlights a significant discrepancy between the optimized dose and the final dose but shows a minimal discrepancy between the full dose and the final dose. This observation suggests that the approximation of the dose influence matrix is the major source of the discrepancies. Therefore, the remainder of this section will primarily focus on addressing the discrepancies arising from the matrix approximation.

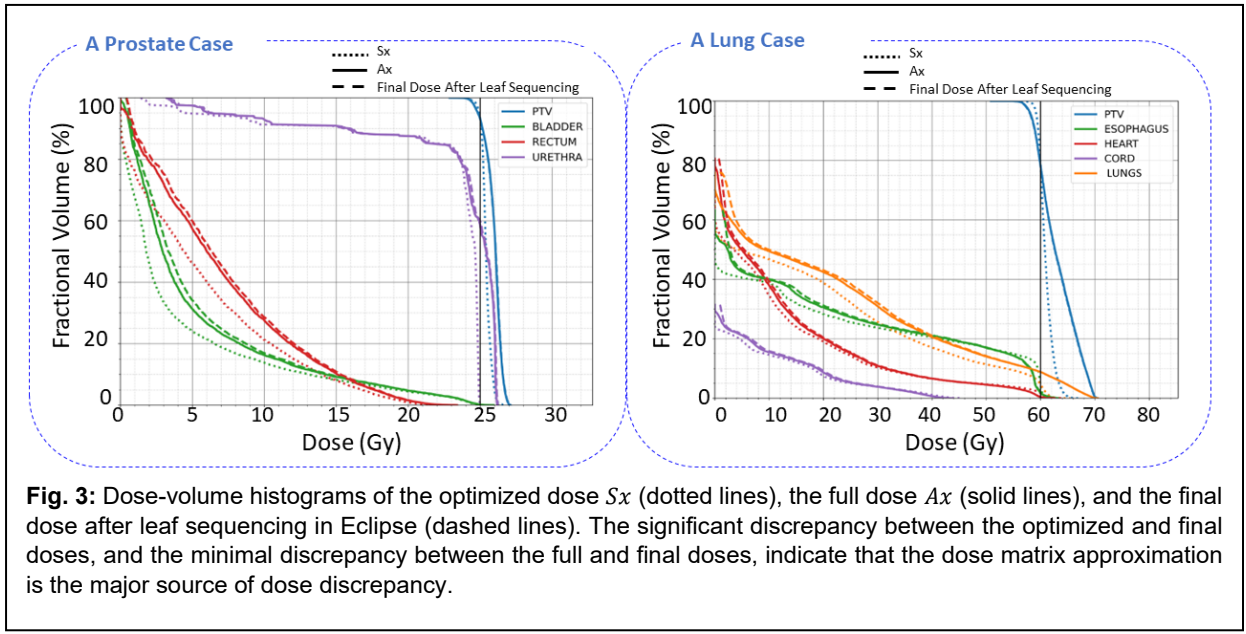

**Fig. 3:** Dose-volume histograms of the optimized dose $Sx$ (dotted lines), the full dose $Ax$ (solid lines), and the final dose after leaf sequencing in Eclipse (dashed lines). The significant discrepancy between the optimized and final doses, and the minimal discrepancy between the full and final doses, indicate that the dose matrix approximation is the major source of dose discrepancy.

Fig. 4 demonstrates the impact of adding a low-rank component $HW$ to the sparse matrix $S$ in the planning optimization process, aiming to minimize the discrepancy between the optimized dose (dotted lines) and the full dose (solid lines). In this experiment, a sparsity threshold $\eta = 1\%$ and a rank $r = 5$ were applied. The figure reveals that adding only rank-5 approximation of the residual matrix $L = A - S$, which consists of small-value dose components, significantly reduces the dose discrepancy. This substantial improvement is attributed to the very low-rank structure of the residual matrix $L$.

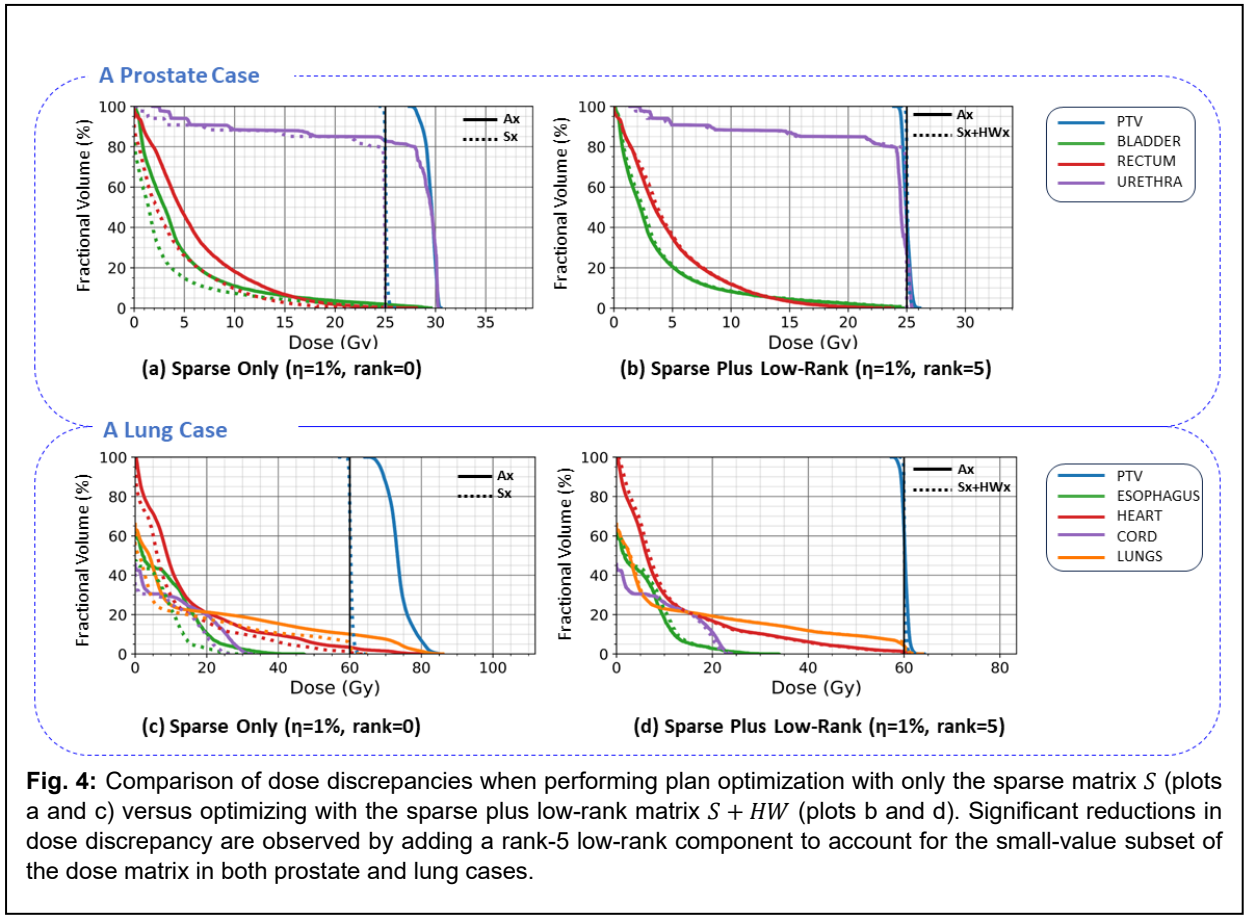

**Fig. 4**: Comparison of dose discrepancies when performing plan optimization with only the sparse matrix $S$ (plots a and c) versus optimizing with the sparse plus low-rank matrix $S + HW$ (plots b and d). Significant reductions in dose discrepancy are observed by adding a rank-5 low-rank component to account for the small-value subset of the dose matrix in both prostate and lung cases.

Fig. 5 explores the effects of two hyperparameters in Algorithm 1, rank and sparsification threshold $\eta$, on the performance of treatment planning optimization by solving Problem (2) for various choices of $\eta$ (1%, 2%, 3%) and rank (0, 5, 10, 20). To assess model accuracy, we quantify the relative dose discrepancy as a percentage: $100 \frac{\|\Delta\|_1}{\|Ax\|_1} = 100 \frac{\|Ax-(Sx+HWx)\|_1}{\|Ax\|_1}$. Computational performance is evaluated by the relative total number of non-zero elements in the sparse and low-rank matrices: $100 \frac{NNZ(S)+NNZ(H)+NNZ(W)}{NNZ(A)} = 100 \frac{NNZ(S)+rank\times(m+n)}{m\times n}$. The number of non-zero elements dictates memory usage and correlates strongly with computational time. However, as demonstrated in Figs. 5.b and 5.d, this correlation is not exact due to the iterative nature of the optimization algorithms and the associated numerical errors and instabilities. Figs. 5.a and 5.c illustrate the trade-offs between accuracy, represented by the relative dose discrepancy, and computational performance, depicted by the relative number of non-zero elements, for various sparsification thresholds and ranks. Points marked with a star ($*$) symbol—indicating a rank of 0—are essentially the sparse-only options. It is evident that these are inferior non-Pareto choices in terms of the trade-offs they offer. For example, the blue star points corresponding to a sparsification threshold of $\eta=1\%$ and rank=0 are outperformed by the orange square points for $\eta=2\%$ and rank=5, suggesting that the compressed representation not only enhances accuracy (i.e., lower dose discrepancy) but also boosts computational performance (i.e., fewer non-zero elements) compared to the sparse-only option. This is further elaborated in Fig. 6. Additionally, Figs. 5.a and 5.c show diminishing returns after adding 5–10 ranks, suggesting that a threshold rank value between 5 and 10 provides an optimal balance between accuracy and performance.

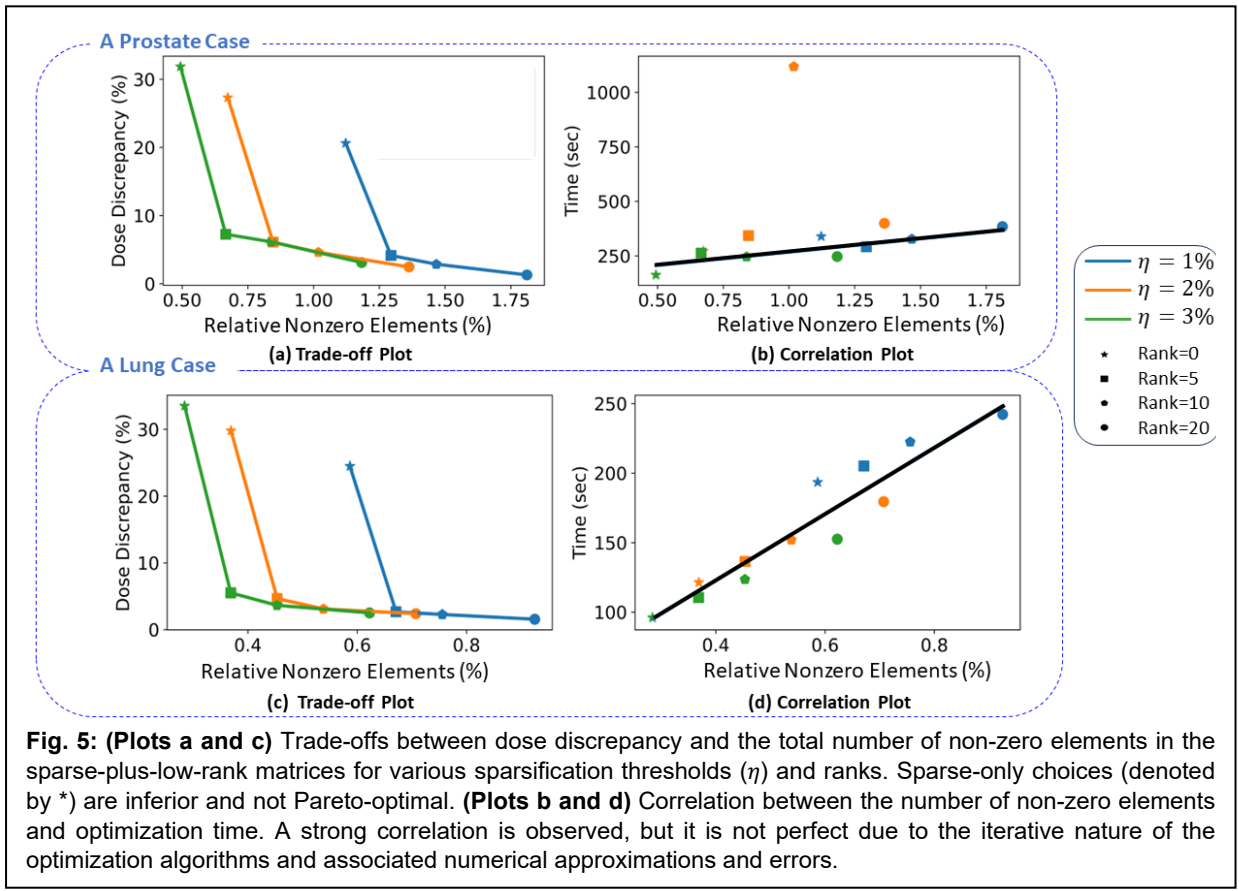

**Fig. 5: (Plots a and c)** Trade-offs between dose discrepancy and the total number of non-zero elements in the sparse-plus-low-rank matrices for various sparsification thresholds ($\eta$) and ranks. Sparse-only choices (denoted by *) are inferior and not Pareto-optimal. **(Plots b and d)** Correlation between the number of non-zero elements and optimization time. A strong correlation is observed, but it is not perfect due to the iterative nature of the optimization algorithms and associated numerical approximations and errors.

Fig. 6 elaborates on the benefits of compression, highlighting improvements in both computational efficiency and accuracy. For accuracy, Fig. 6a to 6d display dose discrepancy comparisons on DVH plots for plans without compression ($\eta = 1\%$ and rank = 0) and with compression ($\eta = 2\%$ and rank = 5), illustrating a clear reduction in dose discrepancies due to compression. Fig. 6e and 6f quantitatively demonstrate these benefits, showing reductions in both computational time and dose discrepancies across ten prostate and ten lung patients. On average, compression reduced dose discrepancies by 73% for prostate cases and 83% for lung cases. Additionally, it decreased optimization time by an average of 20% for prostate cases and 13% for lung cases.

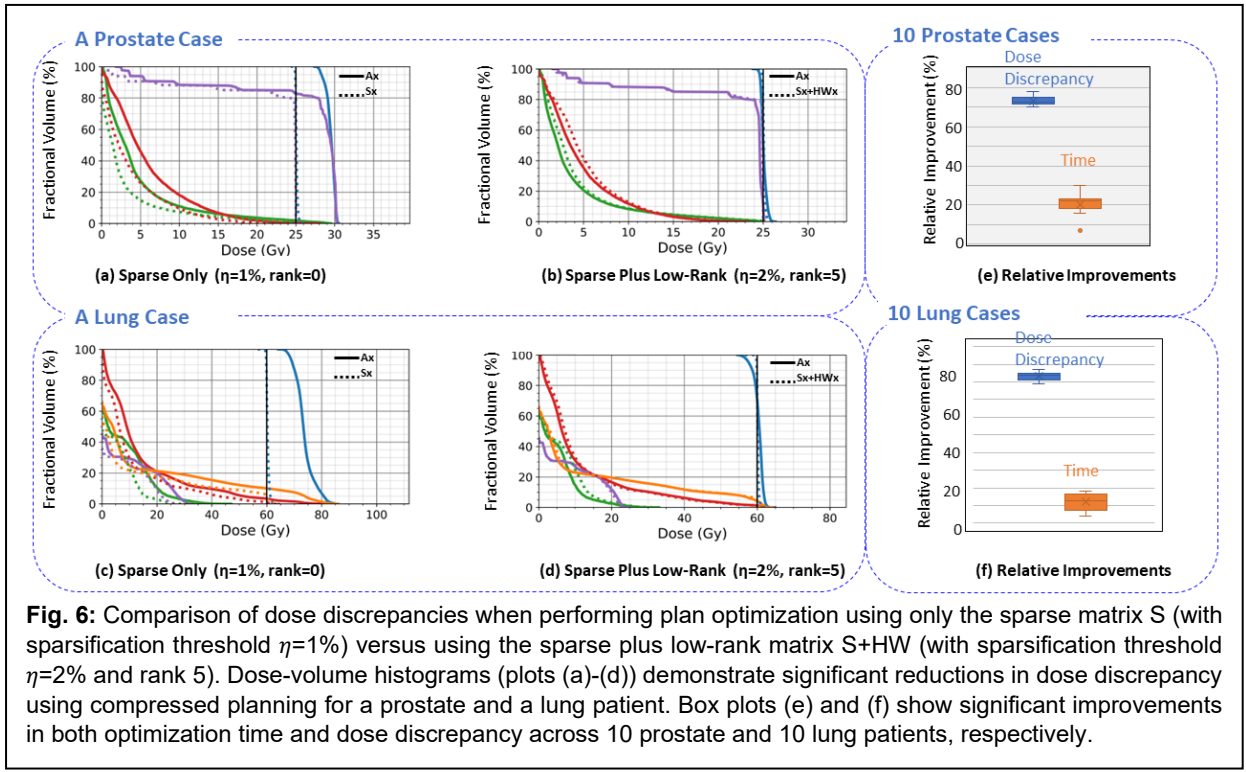

**Fig. 6:** Comparison of dose discrepancies when performing plan optimization using only the sparse matrix S (with sparsification threshold $\eta=1\%$) versus using the sparse plus low-rank matrix S+HW (with sparsification threshold $\eta=2\%$ and rank 5). Dose-volume histograms (plots (a)-(d)) demonstrate significant reductions in dose discrepancy using compressed planning for a prostate and a lung patient. Box plots (e) and (f) show significant improvements in both optimization time and dose discrepancy across 10 prostate and 10 lung patients, respectively.

## 3.2 Correction Loop

Figs. 7 and 8 illustrate the impact of correction steps on reducing dose discrepancies by comparing the DVHs of plans after one and two correction iterations, both with and without compression. The results indicate that although the correction steps are effective, plans without compression require additional iterations. This is particularly significant in constrained optimization settings, as each correction step demands as much computational time as solving the original optimization problem due to the interior point method's lack of warm-start capability[19].

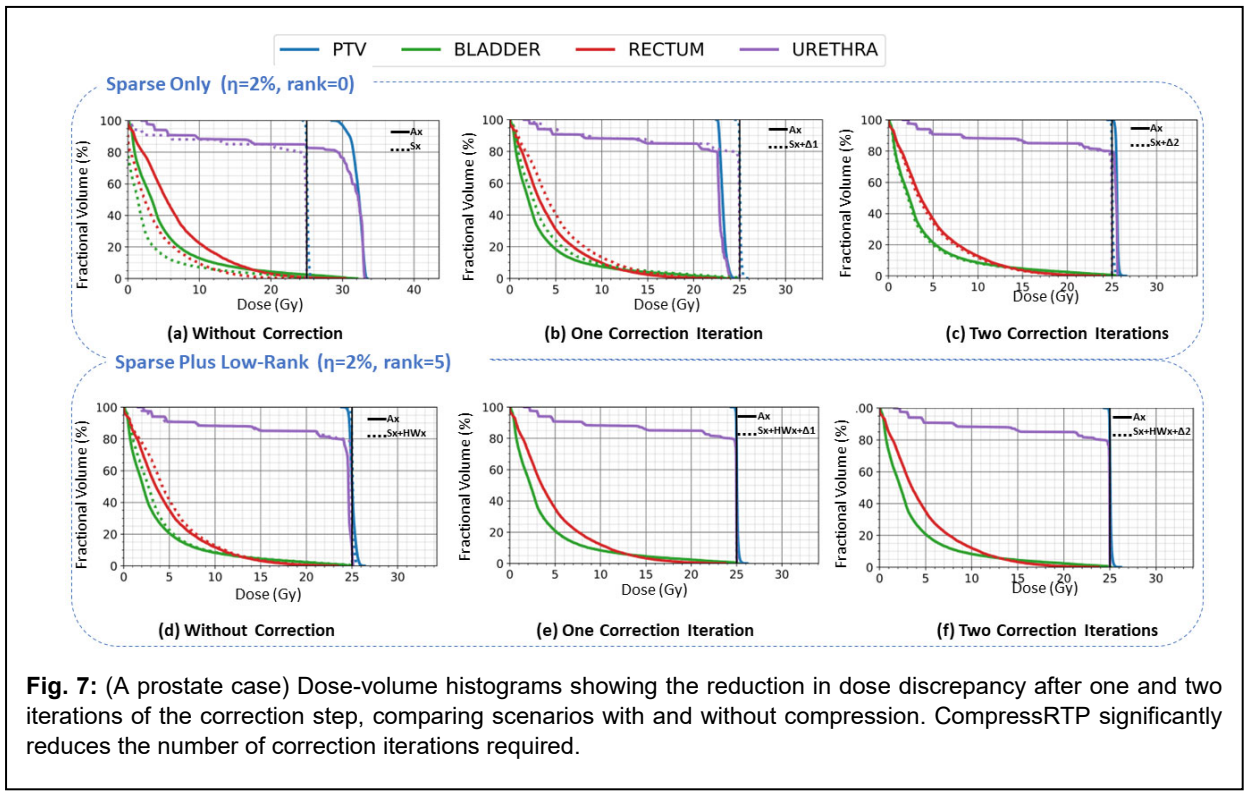

**Fig. 7:** (A prostate case) Dose-volume histograms showing the reduction in dose discrepancy after one and two iterations of the correction step, comparing scenarios with and without compression. CompressRTP significantly reduces the number of correction iterations required.

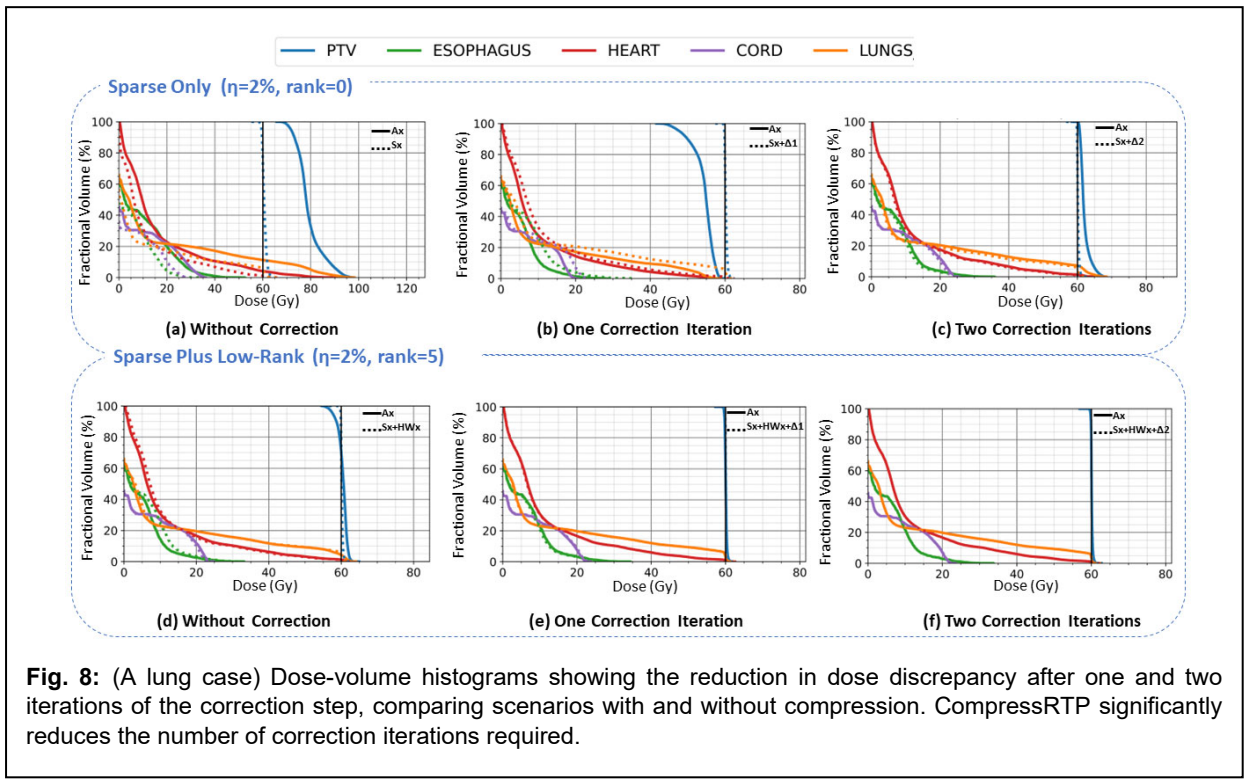

**Fig. 8:** (A lung case) Dose-volume histograms showing the reduction in dose discrepancy after one and two iterations of the correction step, comparing scenarios with and without compression. CompressRTP significantly reduces the number of correction iterations required.

## 3.3 Plan Quality Improvement

Fig. 9 highlights the benefits of the CompressRTP framework in enhancing plan quality for ten prostate patients, evaluated using some clinically relevant metrics. The figure compares plans generated by our in-house automated planning system, ECHO, both with compression (orange boxes) and without compression (blue boxes). All plans were imported into the FDA-approved Eclipse system for final leaf sequencing and dose calculations. To facilitate fair comparison, the plans with and without compression were normalized to have the same $V_{100\%}(PTV)$, typically around 90%. The results in Fig. 9 show that, on average, the CompressRTP framework reduces the maximum doses to the PTV, bladder, rectum, and urethra by 1.5%, 4%, 3.5%, and 3.4%, respectively. Additionally, it improves the mean doses to the rectum and bladder by 12.5% and 8.8%, respectively. Fig. 10 further compares the dose distribution and DVH for a representative prostate case, demonstrating that the benefits of compression are particularly evident in the DVH plots, which show lower doses to the rectum, bladder, and urethra.

Figs. 11 and 12 present similar comparisons for lung patients. Specifically, Fig 11 shows that for ten lung patients, the CompressRTP framework reduces the maximum doses to the PTV, lungs (left and right lungs excluding GTV), spinal cord, esophagus, and heart by an average of 4.4%, 9.5%, 7%, 9.3%, and 8.8%, respectively. Additionally, it improves $D_{95\%}(PTV)$ by 3% and reduces the mean doses to the lungs and heart by 10.8% and 11.2%, respectively. Fig. 12 illustrates the improvements in both PTV coverage and OAR sparing for a representative lung patient.

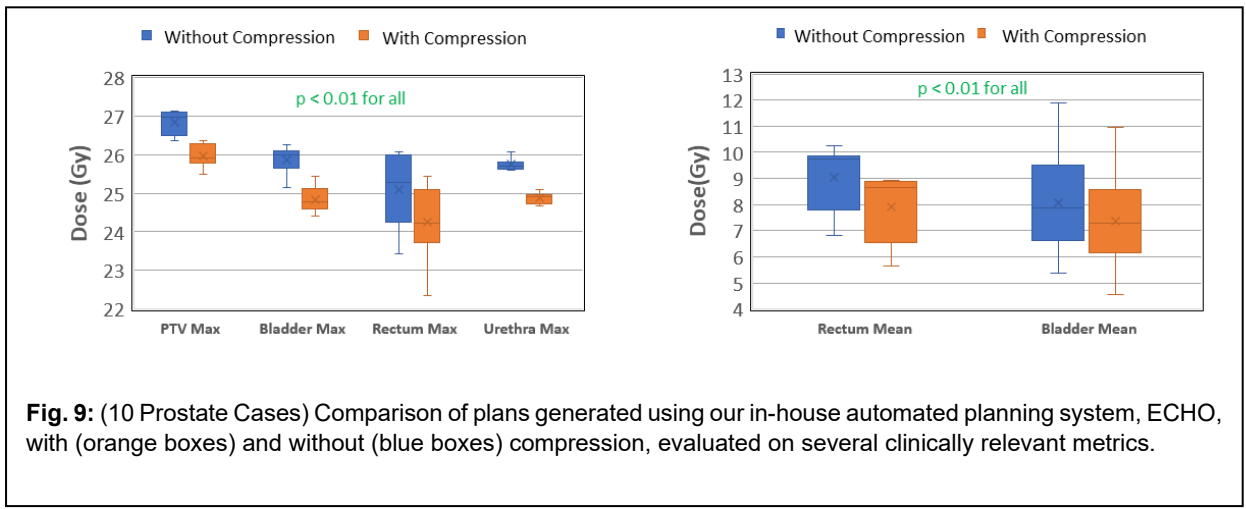

**Fig. 9:** (10 Prostate Cases) Comparison of plans generated using our in-house automated planning system, ECHO, with (orange boxes) and without (blue boxes) compression, evaluated on several clinically relevant metrics.

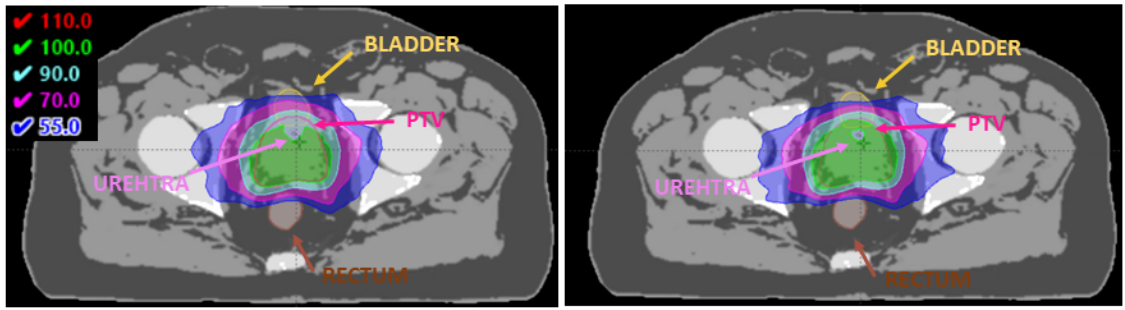

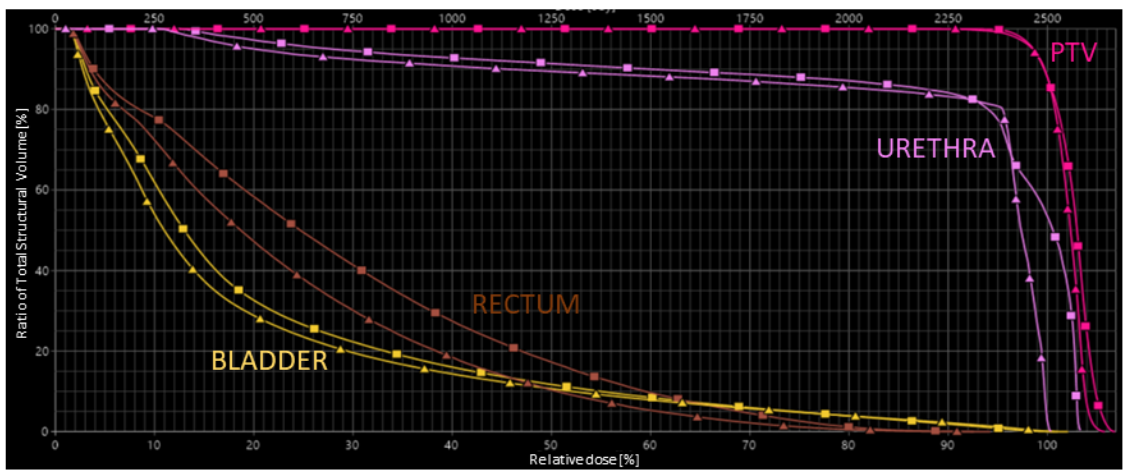

**Fig. 10:** (A Prostate Case) Dose distribution and dose-volume histogram comparisons of automated plans with and without compression.

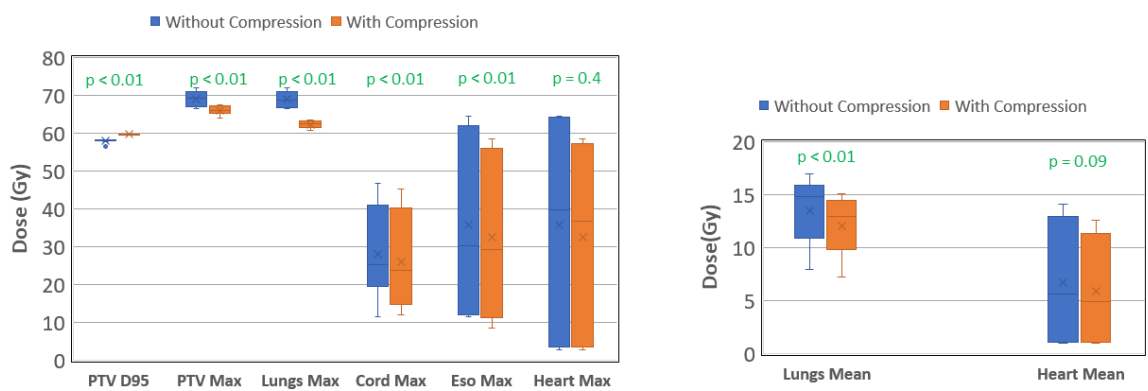

**Fig. 11:** (10 Lung Cases) Comparison of plans generated using our in-house automated planning system, ECHO, with (orange boxes) and without (blue boxes) compression, evaluated on several clinically relevant metrics.

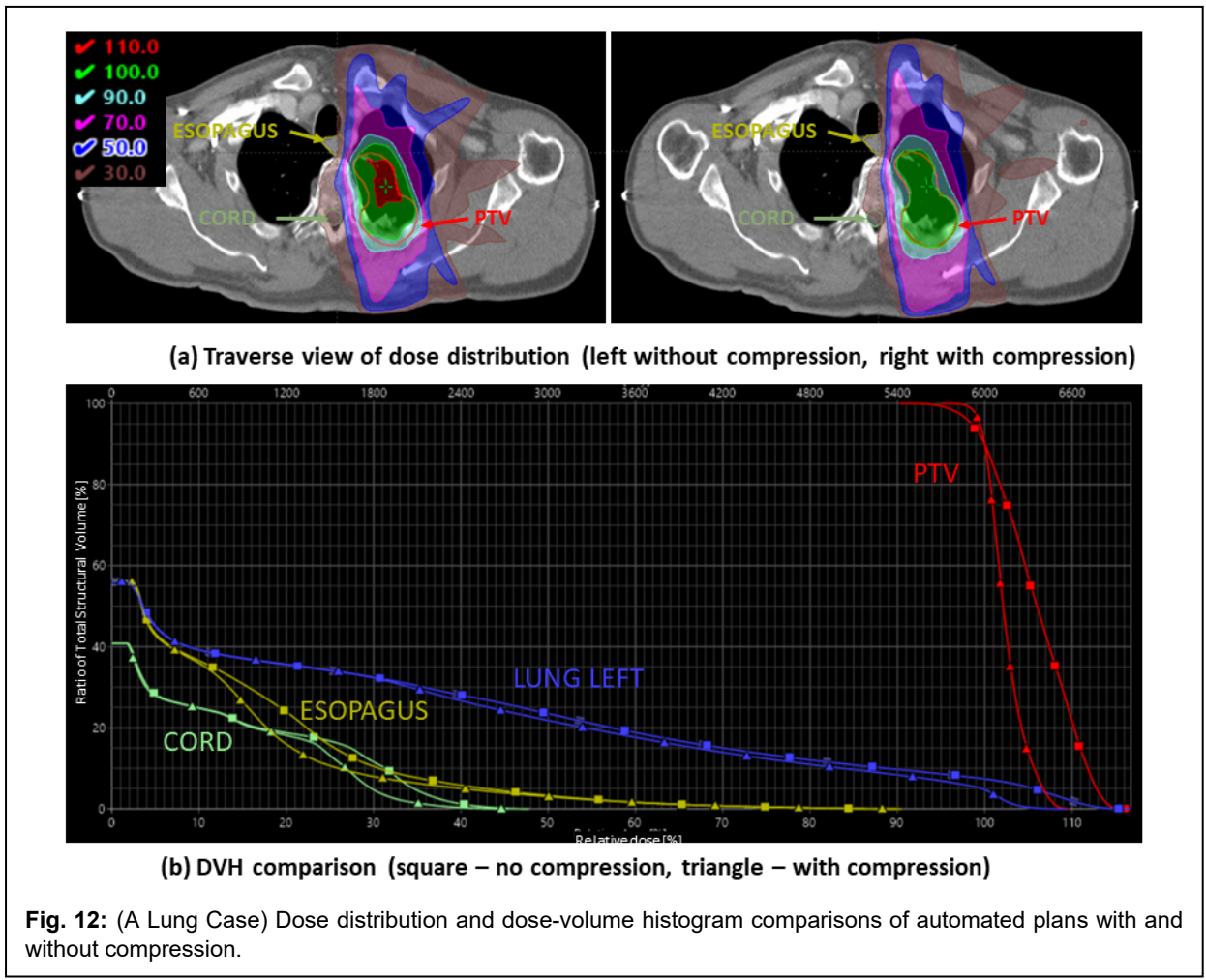

**Fig. 12:** (A Lung Case) Dose distribution and dose-volume histogram comparisons of automated plans with and without compression.

## 3. DISCSUSSION

Significant advances in radiation dose calculation have been made over the past decade, with modern GPU hardware and AI algorithms enabling fast and accurate computations. However, treatment planning optimization—which involves pre-calculating and storing the dose distribution of thousands of beamlets in a dose influence matrix—still relies on less accurate dose calculations. This reliance can lead to sub-optimal treatment plans due to the "garbage-in, garbage-out" phenomenon. On the other hand, when these large dose influence matrices are calculated accurately, they become dense, rendering the optimization process computationally infeasible. In this study, we demonstrate that although these matrices are large and dense, they possess favorable structures and can be highly compressed. This compression allows for an accurate and efficient representation as a sum of sparse and low-rank matrices.

An analogous approach exists in image compression, where a sparse matrix captures the edges of an image, and the smooth regions—due to pixel correlations—form a low-rank structure. Although this technique is rarely used in practice because more advanced methods are available, it offers a useful analogy for dose distribution. We can view the dose distribution as a smooth image with dose gradients forming the edges. This analogy supports our initial motivation: representing the primary and scatter components as sparse and low-rank matrices, respectively, since dose gradients (edges) are primarily shaped by the primary dose contribution. However, accessing the primary and scatter components separately is unnecessary, as the proposed Algorithm 1 efficiently decomposes the matrix without requiring this distinction. Algorithm 1 is computationally efficient, typically requiring only 5%–10% of the time needed to solve the full optimization problem, thanks to modern SVD decomposition algorithms.

Algorithm 1 introduces two hyperparameters: the sparsification threshold ($\eta$), which determines how much dose information is encoded in the sparse matrix, and the rank ($r$), which controls the level of approximation and compression in the remaining low-rank matrix. The choice of these parameters depends on the application (e.g., offline, online, or real-time planning) and the computational framework and available resources. Our results demonstrate that incorporating a low-rank matrix with a rank between 5 and 10 can significantly reduce dose discrepancies and enhance plan quality. For example, as shown in our study, a compressed representation with $\eta = 2\%$ and $r = 5$ outperforms a sparse-only representation with $\eta = 1\%$, both in terms of computational efficiency and plan quality.

Furthermore, we have demonstrated the advantages of the CompressRTP framework by integrating it with our in-house automated planning system, yielding significant improvements for both lung and prostate cases. For ten prostate patients, the framework achieved average mean dose reductions of 8.8% to the bladder and 12.5% to the rectum. For ten lung patients, it resulted in average mean dose reductions of 10.8% to the heart and 11.2% to the lungs.

The "correction loop" has become a standard approach for addressing inaccuracies in dose influence matrix calculations. However, it involves solving multiple optimization problems as correction steps, which can be just as computationally intensive as solving the original problem, particularly in constrained optimization settings. Additionally, previous studies have shown that the solutions obtained through these correction steps do not always converge to the original problem's solution, leading to sub-optimal treatment plans[18]. In the CompressRTP framework, we expect that only one or a few correction steps are required to address inaccuracies arising from lossy compression (since the sparse-plus-low-rank method is not a perfect lossless compression), final leaf sequencing, or uncertainties from AI-based dose influence matrix calculations. However, this area warrants further investigation in future research. Given that the sparse-plus-low-rank decomposition also provides a highly memory-efficient representation of the matrix, it has the potential to enhance implementations on modern GPUs, though this too will require additional research.

It is important to acknowledge that the sparse-plus-low-rank structure is a well-established paradigm, having emerged in various fields such as computer vision, medical imaging, and statistics[37,38]. Historically, this structure has been employed as a form of prior knowledge to recover objects of interest, which manifest themselves either in the sparse or low-rank components. For example, in a surveillance video, the moving objects—typically the focus of analysis—are captured as sparse elements when the video frames are represented as columns in a matrix. However, the application presented in this study represents a novel departure from conventional uses of sparse-plus-low-rank decomposition. Unlike traditional settings where specific components (sparse or low-rank) hold intrinsic importance, our primary goal is not to isolate or interpret these structures but rather to leverage them for computationally efficient matrix representation. In this case, the structure serves purely as a vehicle for optimizing computational efficiency while maintaining data integrity. It is also worth noting that while numerous algorithms exist in the literature for optimally decomposing matrices into sparse and low-rank components, we found these methods prohibitively computationally expensive. In fact, many of them would exceed the computational demands of the optimization problem we aim to solve, defeating the purpose of their implementation in this context. These algorithms, however, remain invaluable in their respective domains where data recovery, not computational speed, is the primary objective.

Finally, although this study has primarily focused on IMRT, our preliminary research suggests that the CompressRTP technique can also be effectively applied to intensity-modulated proton therapy (IMPT). Further investigations are needed to fully assess the benefits of CompressRTP in IMPT and other treatment modalities, particularly those involving larger influence matrices. In such cases, CompressRTP is expected to offer even greater advantages, including applications in beam angle optimization[41,45,46], volumetric modulated arc therapy (VMAT)[3,47], station parameter optimized radiotherapy (SPORT)[14,48], and 4Π[49].

## 4. CONCLUSION

We have developed CompressRTP, a new treatment planning optimization platform that allows us to incorporate accurate dosimetric data—including scattering components—into the optimization process. Combined with recent advancements in AI-based dose calculations and GPU-based optimization frameworks, this platform could enable fast yet accurate treatment planning for emerging online and real-time treatment systems.

## ACKNOLEDGMENTS

This work was partially supported by MSK Cancer Center Support Grant/Core Grant from the NIH (P30 CA008748).

## CONFLICT OF INTEREST

The authors have no relevant conflicts of interest to disclose.

## APPENDIX

Fig. A illustrates that the small-value components of the dose influence matrix (calculated using Eclipse AAA) exhibit a strong low-rank structure, evidenced by the sharp exponential decay of their singular values (red lines). In contrast, the singular values of the large-value components (orange dotted lines) show a much slower decay, closely resembling those of the original matrix—as indicated by the overlap of the solid blue and orange dotted lines. These observations suggest that neither a purely low-rank nor a purely sparse representation is optimal for representing the influence matrix.

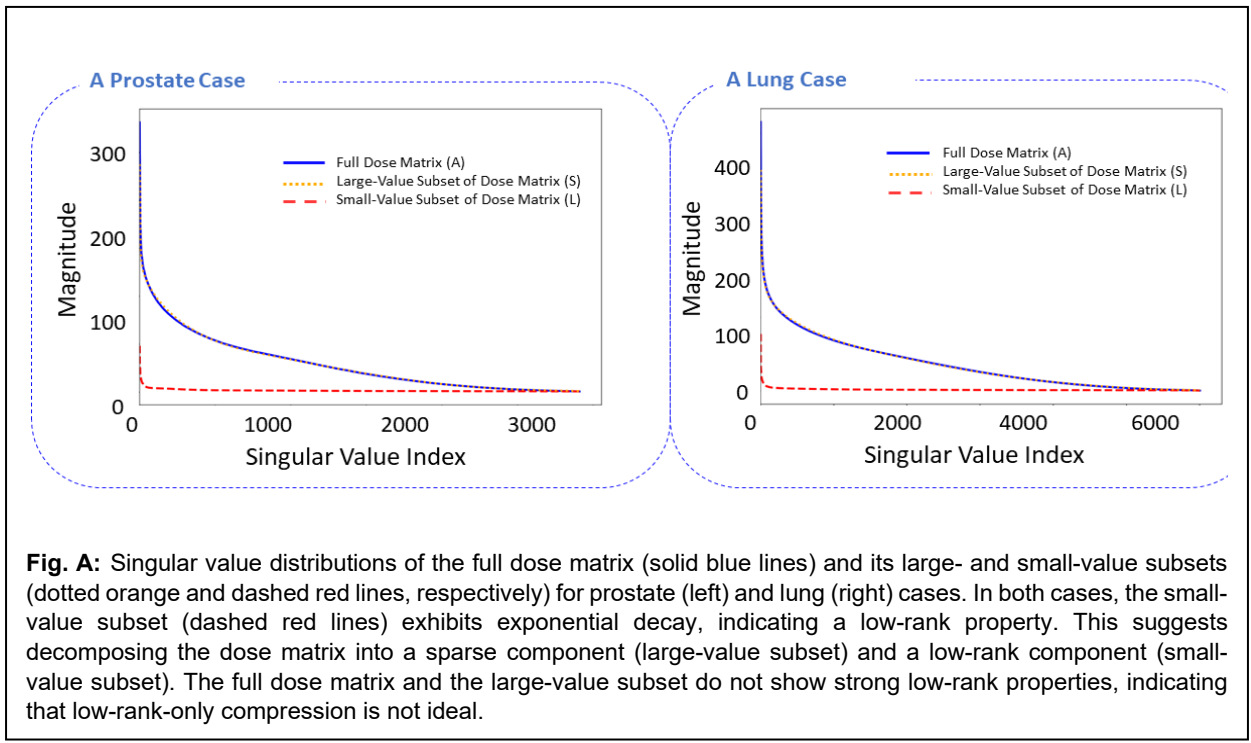

**Fig. A:** Singular value distributions of the full dose matrix (solid blue lines) and its large- and small-value subsets (dotted orange and dashed red lines, respectively) for prostate (left) and lung (right) cases. In both cases, the small-value subset (dashed red lines) exhibits exponential decay, indicating a low-rank property. This suggests decomposing the dose matrix into a sparse component (large-value subset) and a low-rank component (small-value subset). The full dose matrix and the large-value subset do not show strong low-rank properties, indicating that low-rank-only compression is not ideal.